\documentclass[12pt,twoside]{article}
\usepackage{a4,jgg,graphicx}
\begin{document}
\begin{JGGarticle}
        {Geometric simulation of locally optimal tool paths in three-axis milling}
        {M.\ Szilv\'asi-Nagy, Gy.\ M\'aty\'asi, Sz. B\'ela: 
         Geometric simulation of locally optimal tool paths }
        {M\'arta Szilv\'asi-Nagy, Gyula M\'aty\'asi, Szilvia B\'ela}
        {\JGGaddress{Department of Geometry, \\ Budapest University of Technology and Economics\\
                email: szilvasi\at math.bme.hu}
         \JGGaddress{Department of Manufacturing Science and Technology,\\
		Budapest University of Technology and Economics\\
		email: matyasi\at manuf.bme.hu}
	\JGGaddress{Department of Geometry, \\ Budapest University of Technology and Economics\\
                email: belus\at math.bme.hu}
	}
\begin{JGGabstract}
        The most important aim in tool path generation methods is to increase the machining efficiency by minimizing the total length of tool paths while the error is kept under a prescribed tolerance. This can be achieved by determining the moving direction of the cutting tool such that the machined stripe is the widest. From a technical point of view it is recommended that the angle between the tool axis and the surface normal does not change too much along the tool path in order to ensure even abrasion of the tool.  In this paper a mathematical method for tool path generation in 3-axis milling is presented, which considers these requirements by combining the features of isophotic curves and principal curvatures. It calculates the proposed moving direction of the tool at each point of the surface. The proposed direction depends  on the measurement of the tool and on the curvature values of the surface. For triangulated surfaces a new local offset computation method is presented, which is suitable also for 
        detecting tool collision with the target surface and self intersection in the offset mesh.
        \\[1mm]{\em Key Words:} tool path simulation, isophotes,  offset, triangular mesh\\
65D17, 68U20
\end{JGGabstract}

\section{Introduction}

This paper deals with three-axis milling, and presents computations on triangulated and analytical surfaces for determining the optimal moving direction of the tool at a given point of the surface. Basic requirements, such as the tool does not remove material from the target surface and the error remains under a prescribed tolerance, can be fulfilled by keeping the processed surface between the target (part) surface and its outer parallel (offset) surface at distance of a given tolerance. In the focus of this investigations are two problems of tool path planning; how to compute the offset of the mesh in a neighborhood of the actual point, and how to determine the optimal moving direction at any point of the surface represented by a triangular mesh.

In a short survey mathematical approaches are mentioned dealing with geometrical factors in surface milling.
The most frequently used tool path generation methods cut the surface by parallel driving planes in equal intervals or extract isoparametric curves in equally spaced parametric steps.  If the surface has regions of different curvatures, both methods lead to uneven distances between tool paths, consequently, to fluctuating errors. The machining error on the processed surface is measured by the height of the cusps (scallops) left between two adjacent tool paths (Fig. \ref{fig1}). Constant scallop height tool path generation methods for ball-end milling of analytical surfaces are described in \cite{Yoon}, \cite{Feng} and \cite{Kim}, however, the required computations are expensive, and some technical questions (i.e. self-intersections) are not solved in all cases. For triangular meshes a constant scallop height method is presented in \cite{KimYang} for computing the distances between slicing planes in three-axis milling with a ball-end cutter.

Improvements of the frequently used isoparametric tool path generation method is suggested by including additional parameter curves in regions where the error is large \cite{Elber}, or by reparametrizing the surface in order to generate boundary-conform tool paths \cite{Oulee}. The boundary is followed by means of proximity maps in \cite{Held}. The application of two different tool path generation methods (using parallel driving planes or z-level contour parallel curves) is suggested by splitting the surface into ``flat'' and ``steep'' regions according to the angle between the tool axis and the surface normal in \cite{Dong}.

A method for approximating the processed part of the surface around a contact point, for generating the widest stripe along a tool path and for investigating local and global millability with a given cutting tool is presented in \cite{Glaeser} and \cite{Wallner}. The condition for collision-free manufacturing is expressed in terms of the curvature indicatrices of the part surface and the tool surface at their touching point. This local millability can be extended to global millability along an isophotic line. The angle between a given direction (the tool axis) and the surface normal does not change along an isophote, consequently the area of the contact surface of the tool is constant, and the abrasion of the tool is even.
Isophotes are applied to partition the surface into regions, in which a generic side step (the distance between two neighboring tool paths) can be computed from the prescribed maximal scallop height \cite{Ding}. Such regions are computed for detection of interference between the surface and the tool (called usually gouging) in \cite{Yang}. Isophote based tool path generation is proposed in \cite{Xu} and an application to NC-machining in \cite{Han}. A numerical method for computation of isophotic curves on an analytical surface is presented in \cite{Lang}. Despite of the mentioned advantageous properties of isophotes these curves are not suitable milling paths due to their uneven density  causing uncontrollable scallop heights. In general, each set of surface curves fulfills only some conditions for tool paths, but not all at the same time.

Investigation of surface curvatures at the touching point of the target surface and the cutting tool concludes that the most efficient tool feed direction in three-axis milling, which can minimize the total length of necessary tool paths, is the principal direction of the maximal normal curvature \cite{ChenD}, \cite{ChenV}, \cite{Wallner}. Namely, this ``steepest ascending'' tool path leaves the widest track (processed stripe) on the machined surface, as also shown in \cite{Wallner}. The same conclusion is formulated in \cite{Bedi}, where a torodial shaped cutter is investigated on a four and five axis machine. Vector fields are constructed for the optimal moving direction of the tool in \cite{Makhanov}.  A more flexible strategy for parallel plane milling considering the surface roughness is presented in \cite{Quin} by introducing a fitness function proportional to the feed rate and transversal step. Then the acceptable machining directions are computed by optimizing the fitness function.

In surface offsetting algorithms a difficult problem is to detect and to remove self-intersections. For triangulated surfaces different offsetting methods have been published. Faces, edges and vertices are offset using half cylinders and spheres, then gaps are filled with blending methods and overlaps are removed.  By slicing this corrected offset triangular mesh with a series of planes three-axis tool paths are calculated \cite{KimY}. By a local offsetting scheme facets are translated in the normal direction, convex edges to trimmed cylinders and convex vertices to trimmed spheres. Concave edges are removed from offsetting. Then in tool path generation the intersection with a driving plane results in circle arcs and line segments which are sorted and linked, overlapping portions are removed in order to generate a compound curve for cutting path \cite{Jun}, \cite{KimDS}.  In a z-level contour tool path generation algorithm the triangulated surface is sectioned by a horizontal drive plane, then at each vertex 
of the section polygon an offset vector is created. A tool path is generated sequentially along the section polygon while checking and removing intersections and using line fillets to connect offset segments \cite{Chuang}. Each triangle is offset in \cite{Park}, the sliced segments in the drive plane are swept and chained into monotone polygonal tool paths. A different offsetting method is presented in this paper in Section 2.

In Section 3 tool paths on triangulated surfaces are simulated by moving the tool in principal directions. In Section 4 proposals for optimal moving direction at a given point  and its computation on analytical surfaces are presented.

\section{Mesh offsetting and computing the processed part of the surface at a point}

The result of the milling process with ball-end or toroidal tools is an approximation to the target surfaces, where the shape shows scallops (cusps). The maximal height of these scallops has to be kept within a given tolerance, i.e. the machined surface must lie between the target (part) surface and its offset by the given tolerance (Fig. \ref{fig1}). The piece of the tool surface lying in this region will be named contact surface of the tool. The curve of intersection of the tool surface and the offset surface is the boundary of this contact surface. The processed patch around a given touching point of the tool and the target surface is bounded by the projection of this boundary curve onto the part surface. The set of all processed patches should cover the entire part surface.  In Fig. \ref{fig2} a smooth surface, its offset, a touching ball-end tool and the processed patch on the surface are shown.

\begin{figure}
\psone{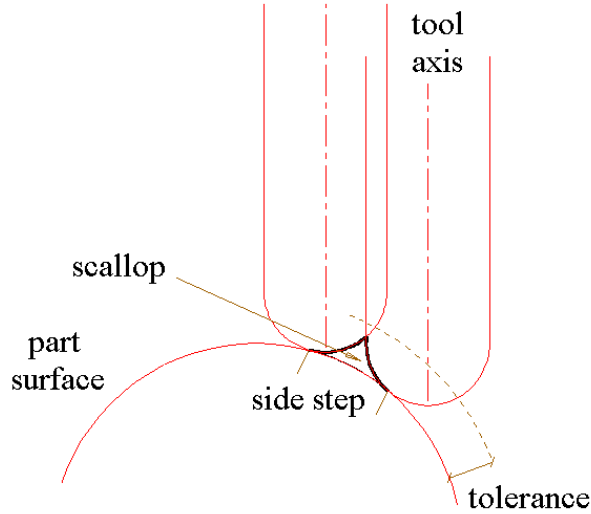}{8cm}
\caption{Milling with given tolerance}\label{fig1}
\end{figure}

\begin{figure}
\psone{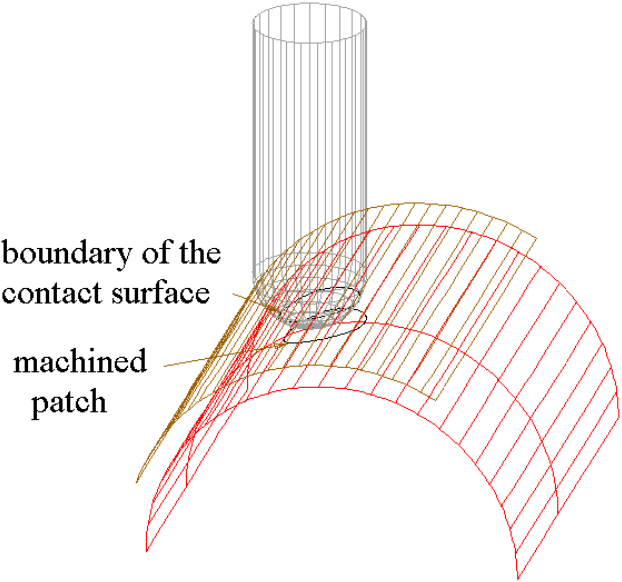}{9cm}
\caption{The processed patch on the part surface}\label{fig2}
\end{figure}

In order to compute the processed patch at a given point, first the offset surface has to be generated.

In the case, when the surface is represented by a triangular mesh, the offset will be computed in $n$ normal planes. Each normal plane passes through the touching point ${\rm P}$ of the tool, through the normal vector ${\bf N}$ of the actual triangle facet containing this point and an arbitrary direction in the plane of the triangle. The offsetting is carried out on the segments intersected by the normal plane from the triangle mesh (Fig. \ref{fig3a}). If the $i$th plane intersects the $j$th triangle in the segment ${\rm Q}_{1j},\, {\rm Q}_{2j}$, its offset is ${\bf Q}_{1j}+\epsilon ^{\prime}{\bf N}_j,\, {\bf Q}_{2j}+\epsilon ^{\prime}{\bf N}_j$. Here $\epsilon$ is the given tolerance, ${\bf N}_j$ is the normal vector of the $j$th triangle and
$\epsilon ^{\prime}=\epsilon \cos \alpha$ ($\alpha $ is the inclination angle between the normal vector ${\bf N}_j$ of the intersected $j$th facet and the $i$th section plane. $\epsilon$ is shown by a thick segment and $\alpha$ is denoted by one arc in Fig. \ref{fig3a}). The position vector of a point is denoted by the corresponding boldface letter. In a convex region gaps arise between the offset segments, but in a concave region the adjacent segments intersect. Filling the gaps and removing self-intersections of the offset mesh are carried out by two-dimensional methods in each section plane. The obtained polygonal line intersects the normal section of the tool end in two points, ${\rm B}_{1i}$ and ${\rm B}_{2i}$ which are lying on the boundary curve of the contact surface of the tool and the offset mesh (Fig. \ref{fig3b}).
The projections of ${\rm B}_{1i}$ and ${\rm B}_{2i}$ onto the mesh give the endpoints of a diameter of the processed patch.  This computation in the normal planes ($i=1,\dots ,n$) results in $2n$ points lying on the boundary curve of the processed patch.
In Fig. \ref{fig4} the part of the offset surface within the ball-end of the tool is shown by 24 diameters above the base triangle. The boundary of the processed patch is drawn by a polygonal line on the mesh. Additional section planes can be included easily, if necessary.

\begin{figure}[h!]
\psone{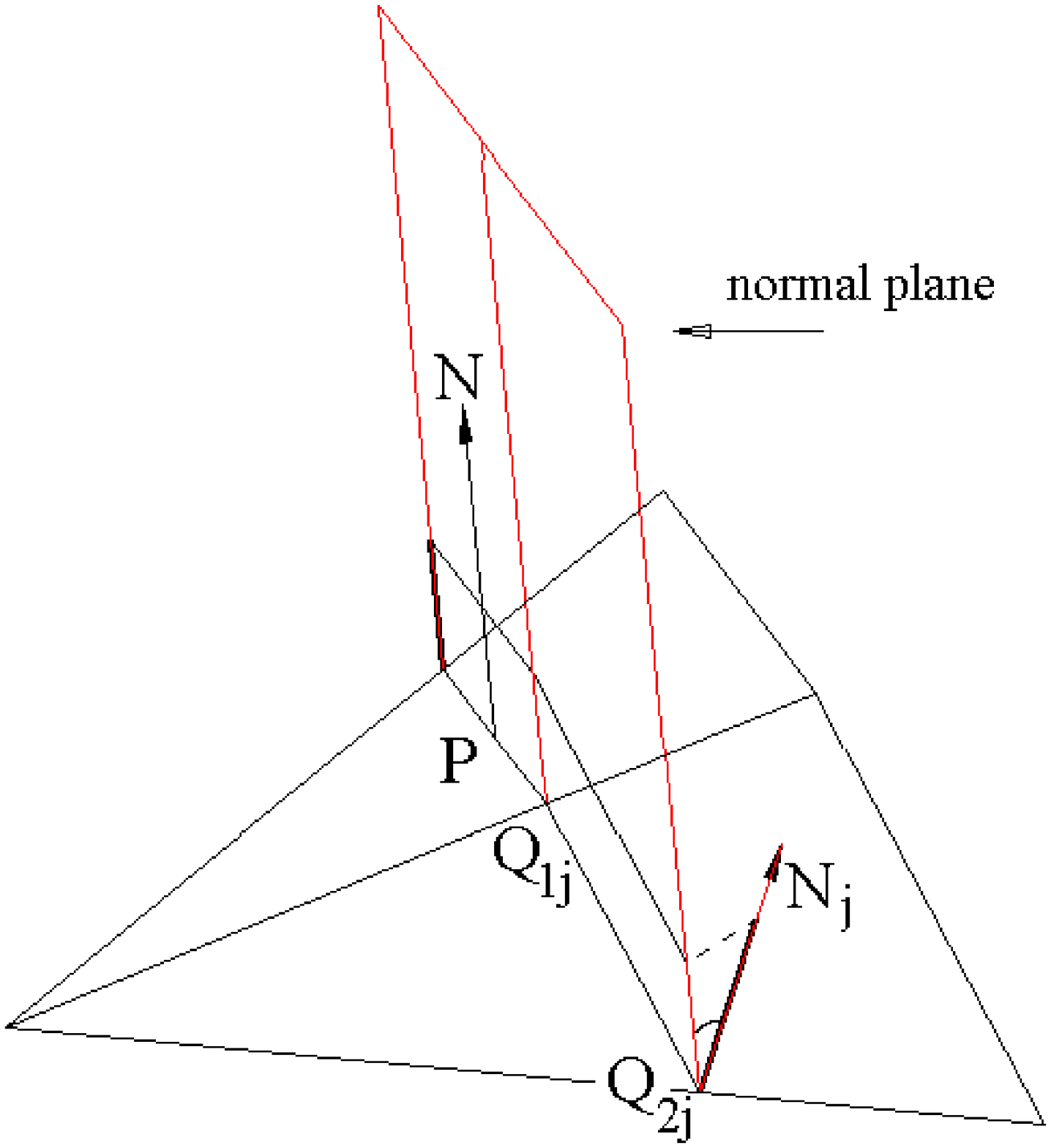}{6.5cm}
\caption{Offset construction in a normal plane}\label{fig3a}
\psone{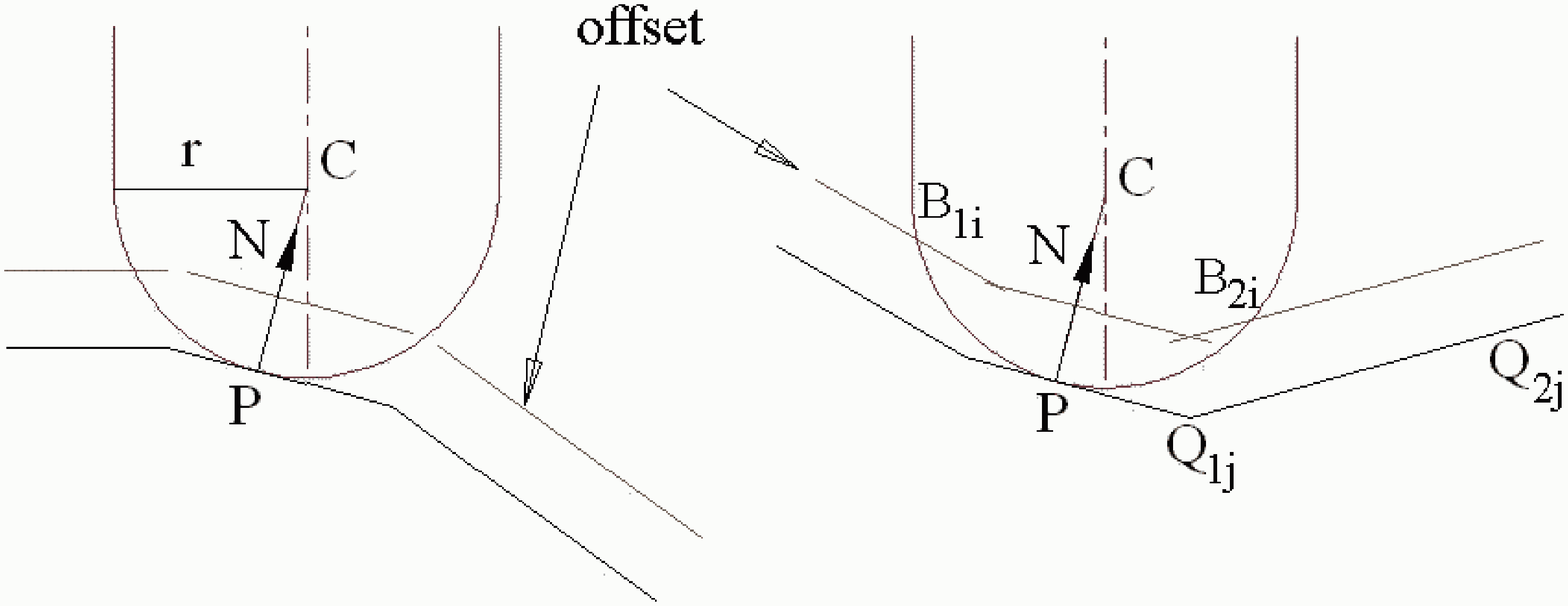}{10cm}
\caption{Computation of the intersection with the tool in a convex and a concave region}\label{fig3b}
\end{figure}

\begin{figure}
\psone{millfig4}{5cm}
\caption{Part of the computed offset within the ball-end and the processed patch on the mesh of a quadratic surface. The actual triangle is lined.}\label{fig4}
\end{figure}

This local offsetting method is novel in the literature. It has the advantage that only two dimensional algorithms are necessary in the computation. Self intersection in the offset mesh and tool collision with the target surface can be detected by simple computation. Moreover, registering the directions of the normal planes, where such problems arise, a method for partitioning the surface could be developed for different milling tools and strategies. The applied edge-oriented polyhedral data structure defined on the mesh is very effective in determining plane sections of the mesh \cite{Szilvasi}.

\section{Moving the tool in principal direction}

First, the requirement is considered that the machined stripe along a tool path is the widest. Such tool paths will usually minimize the total length of milling paths \cite{ChenD}, \cite{ChenV}. The processed patch by the tool around the touching point is determined by its $n$ diameters lying on the surface.  Now the tool will be moved in the direction perpendicular to the largest diameter. The next touching point of the tool will be chosen in this direction on the boundary of the processed patch. The suggested moving direction  is similar  in \cite{Wallner} in a different computational approach using Dupin-indicatrices.

In Fig. \ref{fig5} two tool paths on a quadratic saddle surface are shown with moving directions producing the widest stripe. The side step of the tool is chosen in such a way that the series of the processed patches overlap.

\begin{figure}[h]
\psone{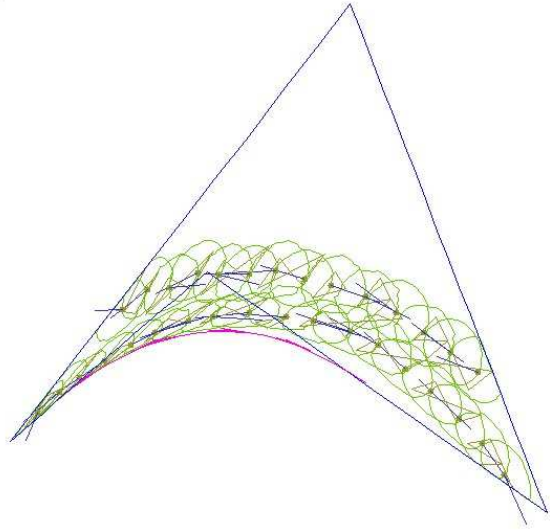}{8cm}
\caption{Tool paths on the  quadratic saddle surface shown in Fig. \ref{fig4}}\label{fig5}
\end{figure}

Tool paths of the widest machined stripe are shown on a cylindrical surface in Fig. \ref{fig6} and \ref{fig7}  A floating patch above one base triangle is also shown which is the part of the offset mesh within the ball-end of the tool. The moving directions are shown by straight line segments. Each is emanating from the touching point on the actual triangle.  The diameters of the processed patches are the shortest in the directions perpendicular to the generators in the case, when the tool touches the cylinder on the convex side, and they are parallel to the generators, when the tool is on the concave side of the surface. In the case of these ``synthetic'' meshes of a cylindrical surface the largest and shortest diameters of a processed patch are perpendicular to each other, and they are practically the principal directions. Investigations have been made on different synthetic meshes which have shown that the direction of the widest processed patch is very close to the principal direction of the largest curvature computed in 
a sufficiently large region around the contact point. For the estimation of the principal directions at a point of the triangulated surface a method is described in \cite{Szilva1} and \cite{Szilva2} by lying a circular disc onto the mesh and computing the minimal and maximal chord lengths between the end points of its curved diagonals. In this concept the curvature values are ordered to the triangle facets of the mesh, and approximated surface normals are replaced by facet normals.

The applied normal curvature estimation method  constructed for triangular meshes is suitable for the computation on meshes also in such cases, where the known vertex oriented computations do not work. Namely, in the mesh of the shown cylinder all the vertices are situated on the surface boundary. To the contrary, frequently used vertex oriented methods require good vertices in the inside of the actual region.

\begin{figure}
\psone{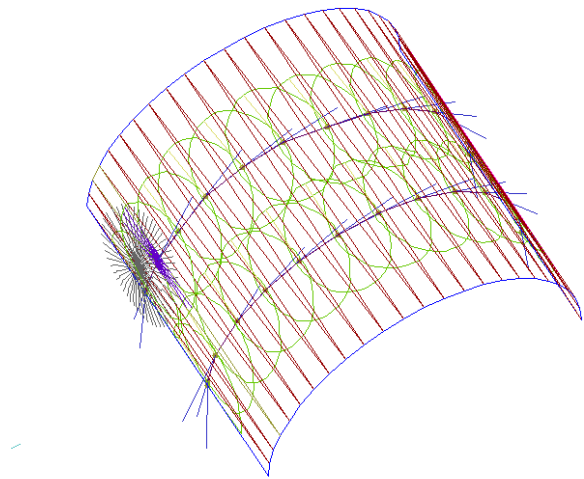}{8cm}
\caption{Tool paths on a cylindrical surface machined from the convex side}\label{fig6}
\end{figure}

\begin{figure}
\psone{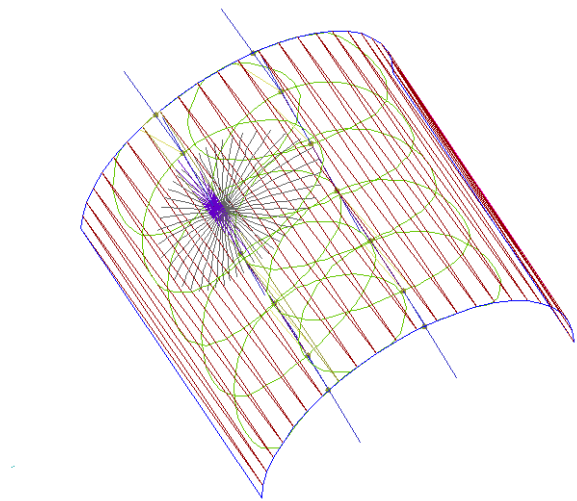}{8cm}
\caption{Tool paths on the same cylindrical surface machined from the concave side}\label{fig7}
\end{figure}

Even if  the total tool path computed in the widest stripe direction can be the shortest, this kind of curves are not always appropriate tool paths due to technical reasons. Namely, the angle between the tool axis and the surface normal changes to much on a curved surface, which leads to uneven abrasion of the tool. The computed moving direction could be  combined with other tool path generation algorithms as suggested also in \cite{Wallner}.

\section{Locally optimal moving direction and its computation on analytical surfaces}

Now a technical requirement will be considered that the abrasion of the tool  is even during the milling process. This requirement is fulfilled, if the angle between the tool axis and the surface normal is constant along a tool path. Such curves are the isophotes on smooth surfaces. (The investigation of other material and kinematical factors is not the object of this paper.)

As the isophotes are not suitable tool paths, a compromise will be suggested, and the tool will be moved from the contact point neither in the direction of the widest processed patch, nor along an isophote. A first idea was to choose the bisector of these two directions. The computation has been made on an analytical surface given by the function $f(x,y)$, assuming that the piece of the surface to be machined is visible from the tool axis (i.e. the $z$-axis) direction \cite{Szibema}. In such cases the $z$-projection method can be applied, when the processed patch is computed by projecting the boundary of the contact surface parallel to the $z$-direction, and the step size is given in the $xy$-plane \cite{Chuang}.
For the points ${\rm B}(x,y,f(x,y))$ of the boundary curve of the contact surface the equation
$$
r={\rm distance}({\rm C},{\bf B}+\epsilon {\bf N}(x,y)), \quad
{\bf N}(x,y)=\frac{(f^{\prime}_x(x,y),f^{\prime}_z(x,y),-1)}{\sqrt{f^{\prime \, 2}_x(x,y)+f^{\prime \, 2}_y(x,y)+1}}
$$
yield, where $r$ is the radius, ${\rm C}$ is the center point of the ball-end of the tool, and $\epsilon$ is the offset distance. $\langle ,\rangle $ denotes the dot product.
The points of the isophote passing through the point ${\rm P}$ satisfy the equation
$$
\cos ^{-1}\langle {\bf N}_p,{\bf a}\rangle = \cos ^{-1}\langle {\bf N}(x,y),{\bf a}\rangle \, ,
$$
where ${\bf a}$ is the unit vector of the tool axis direction. In the specific coordinate system, where ${\bf a} || z$, this equation is equivalent to
$$
f^{\prime \, 2}_x(x_P,y_P)+ f^{\prime \, 2}_y(x_P,y_P)=
f^{\prime \, 2}_x(x,y)+ f^{\prime \, 2}_y(x,y) \, .
$$
Both equations can be solved numerically resulting in a finite number of points lying on the required surface curves, i.e. on the boundary of the processed patch and on the isophote, respectively.

Fig. \ref{fig8} shows on a cubic surface how the processed patch around a contact point and the isophote through this point are situated.  From the two common points of the curves two bisector directions can be determined (Fig. \ref{fig9}). One of them (which doesn't lead to a zig-zag curve) has been chosen for tool path direction. Few points of three generated tool paths with the processed patches are shown in Fig. \ref{fig10} The step size is constant.
The problem of varying distances between the tool paths occurs here as well, as also in the frequently used methods (\cite{Ding}, \cite{Elber}).

\begin{figure}[t]
\psone{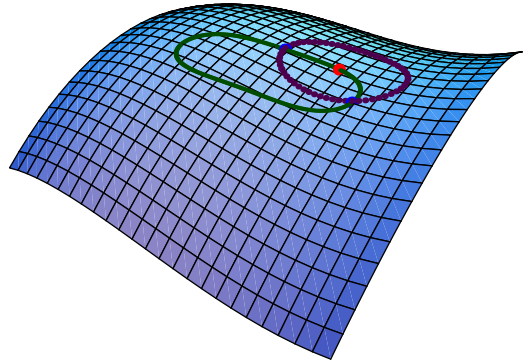}{8cm}
\caption{A processed patch around the contact point of the tool and the isophote through this point on a cubic surface}\label{fig8}
\end{figure}

\begin{figure}
\psone{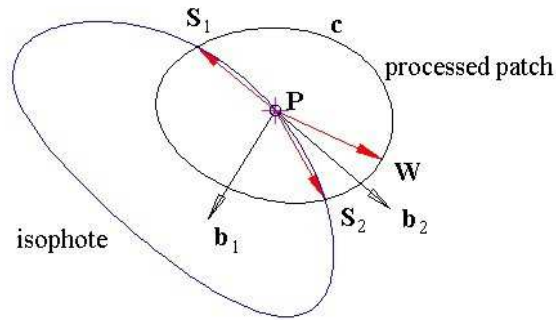}{9cm}
\caption{Bisector directions ${\bf b}_1$ and ${\bf b}_2$ of the angle $\langle {\rm W},{\rm P},{\rm S}_1\rangle$ and $\langle {\rm W},{\rm P},{\rm S}_2\rangle$, respectively.}\label{fig9}
\end{figure}

\begin{figure}
\psone{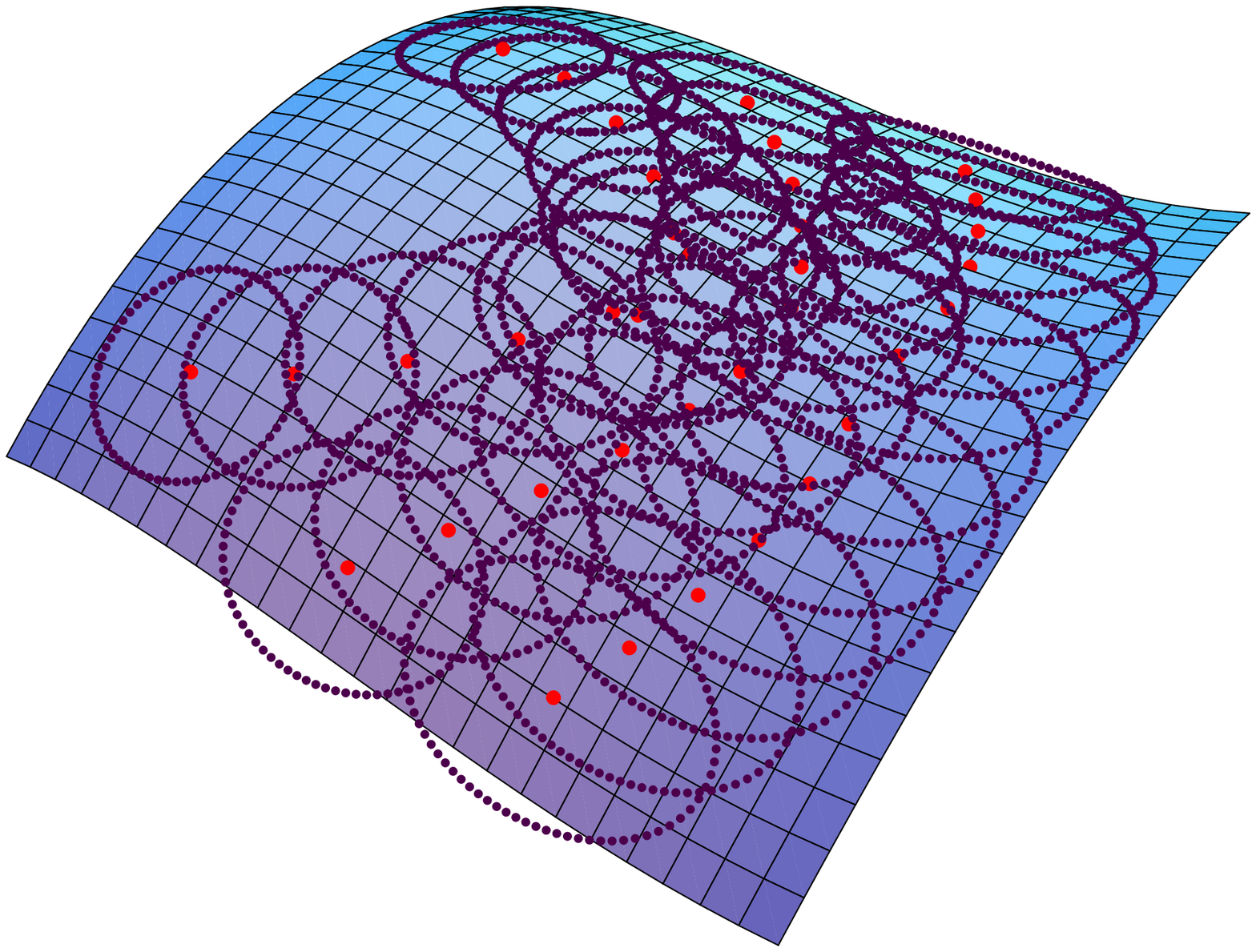}{8cm}
\caption{Tool paths generated in the bisector of the widest stripe and the isophote directions.  To each point the processed patch is shown.}\label{fig10}
\end{figure}

A more reasonable idea for finding the optimal direction at any point of the surface is to consider the change of the angle between the tool axis and the surface normal (i.e. the inclination angle of the isophote passing through the actual point) while the tool is moving into a next position. This will determine the correction factor of the declination from the widest stripe direction. If the surface has small curvature, the surface normal doesn't change very much while moving in the principal direction of the maximal curvature. Therefore, the tool can be kept on this surface curve, because the abrasion will be even. In the case of larger curvature values the tool should be moved rather towards the corresponding isophote direction.

In a generic case the suggested moving direction will be determined between the widest stripe direction and the direction pointing to the isophote point on the boundary of the processed patch ${\bf c}$ in such a way that it divides the arc between these two directions in the ratio $(1-\cos \beta): \cos \beta$. $\beta$ denotes the difference of the inclination angles of the surface normals to the tool axis at the two corresponding points on the processed patch boundary. In Fig. \ref{fig11} the vector ${\bf w}$ is pointing into the direction of the widest stripe and ${\bf s}$ is pointing to the point ${\rm S}$ of the isophote passing through the contact point ${\rm P}$.
The boundary ${\bf c}$ is represented by $2n$ points ${\rm S}_i$ which are the projections of the points lying on the boundary of the contact surface denoted by ${\rm B}_{1i}$ and ${\rm B}_{2i}$ in Fig. \ref{fig3b} ($n$ is the number of the section planes). The isophote point ${\rm S}$ on ${\bf c}$ can be approximated by one of the points ${\rm S}_i$, in which for the angles
$$({\bf N}_{S_i},{\bf a})\angle \approx ({\bf N}_P, {\bf a})\angle $$
yields, computing from the condition
\begin{eqnarray*}
 |\cos ^{-1}\langle {\bf N}_S, {\bf a} \rangle -  \cos ^{-1}\langle {\bf N}_P, {\bf a} \rangle |=
min |\cos ^{-1}\langle {\bf N}_{S_i}, {\bf a} \rangle -  \cos ^{-1}\langle {\bf N}_P, {\bf a} \rangle |&,& \\
{\bf S}_i \in {\bf c}, \, i=1,\dots ,2n &,&
\end{eqnarray*}
where ${\bf N}_{S_i}$ is the normal vector of the triangle containing the point $S_i$.
That means, the direction of ${\bf N}_S$ should fit on the cone determined by the isophote's angle of ${\bf N}_P$ (denoted by single arc in Fig. \ref{11}). The vectors ${\bf N}_S$, ${\bf N}_P$, ${\bf N}_{S_i}$ and ${\bf a}$ are unit vectors.

Let $\beta$ denote the difference of the angles which are formed by the surface normals ${\bf N}_W$ and ${\bf N}_S$ to the axis direction ${\bf a}$.
$$
\beta =|\cos ^{-1}\langle {\bf N}_W, {\bf a}\rangle - \cos ^{-1}\langle {\bf N}_S,{\bf a}\rangle |, \quad \| {\bf N}_W \|=1, \> \| {\bf N}_S\| =1, \, \|{\bf a}\| =1.
$$
The proposed new moving direction
${\bf q}$ is determined by the arc length  of the processed patch boundary between ${\bf w}$ and ${\bf s}$:
$${\rm arc}({\bf q},{\bf w})=(1-\cos \beta )\cdot {\rm arc}({\bf s},{\bf w}).$$
Consequently, a zero, or a very small $\beta $ does not change the moving direction ${\bf w}$. This $\beta$ depends also on the radius of the ball-end. The bigger the tool, the bigger is the processed patch. Therefore, the change of the inclination angle of the normals is larger. We remark that the factors  included in the fitness function introduced in \cite{Quin} have the same effect on the moving strategy.

This path correction does not influence the abrasion of a ball-end tool, it has importance in milling with other cutters. The computation for a toroidal cutter does not differ from the presented computation above, but the demonstrating figures have been made with ball-end tools.

Fig. \ref{fig12} shows a tool path computed by the suggested method. The path in the widest stripe direction (practically a parallel circle of the cylinder) is shown with the processed patches around the computed contact points. The corrected tool path situated lower shows a declination. The radius of the ball-end of the tool is in this case eight times the average size of the triangles. The declination of the new path is less with a smaller radius. In the case, when the radius is only four times the average size of the triangles, the tool will move approximately along the path of the widest stripe. The tolerance was one third of the tool radius in this case, which is the distance of the floating piece of the offset and the part surface (see on the left-hand side in the figure). Note that the tool paths computed on a cylinder surface in a completely different way in \cite{Quin} show a similar shape.

\begin{figure}
\psone{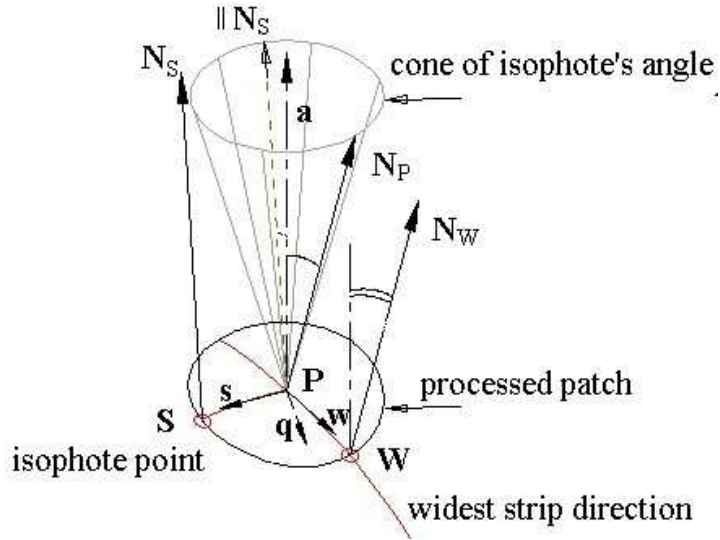}{10cm}
\caption{The proposed moving direction ${\bf q}$ is computed between the widest stripe direction ${\bf w}$ and the isophotic direction ${\bf s}$. }\label{fig11}
\end{figure}

\begin{figure}
\psone{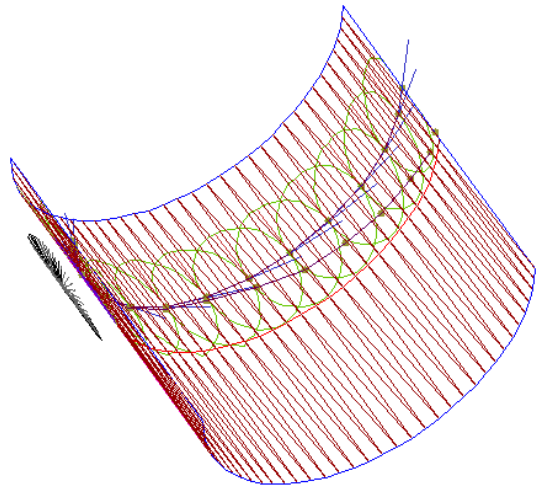}{8cm}
\caption{Tool paths on the cylinder in the widest stripe direction and in the proposed moving direction.}\label{fig12}
\end{figure}

The next examples are computed on parametric surfaces.
Though the surface is presented by a vector function ${\bf r}(u,v)$, $(u,v) \in [u_1,u_2]\times [v_1,v_2]$, which is at least twice differentiable in a sufficiently large neighborhood of ${\rm P}$, the computation is carried out in $n$ normal planes producing $2n$ points of the surface curve ${\bf c}$ bounding the processed patch. In this way, the time consuming and unstable process of computing the projection of a point onto the surface will be replaced by a very effective method.

Let's say, the actual tangent direction vector is ${\bf s}_i$, $i\in \{ 1,\dots  n\}$.
The normal plain containing the required point ${\rm S}_i$ is determined by the vectors ${\bf s}_i$ and the surface normal ${\bf N}_P$. The decomposition of ${\bf s}_i$ in the tangent plane is
$$ {\bf s}_i = a_i \left. \frac{\partial {\bf r}(u,v)}{\partial u} \right|_P
+ b_i \left. \frac{\partial {\bf r}(u,v)}{\partial v} \right|_P \, .$$
Let the point to be projected onto the surface be ${\rm B}_i$ lying in this normal plane. First, $a_i$ and $b_i$ are computed as solutions of the system of linear equations from the decomposition of ${\bf s}_i$. Then series of points are generated in the parameter domain with an appropriate $\Delta t$, which are
$$(u_k,v_k)= (u_P,v_P)+ k \cdot \Delta t \frac {(a_i,b_i)}{\sqrt {a_i^2+b_i^2}}, \quad
k=1,\dots ,K \, .$$
The corresponding points on the surface are determined by the vectors
${\bf r}(u_k,v_k)$. The required point ${\rm S}_i$ is determined by the parameter values $(u_k,v_k)$ for which the distance of the surface point ${\bf r}(u_k,v_k)$ to
${\rm B}_i$ is minimal $(k \in \{1,\dots ,K\}$.

Numerical tests have shown that the results of the presented approximation of the projection onto the surface are satisfying compared to the "real" projections computed by the program package Mathematica by searching for a nearest surface point to the one to be projected without any constraints. The difference between the results of the two methods is within the error bounds of the computation in other procedures of the complete task.

In the Figures \ref{fig13} and \ref{fig14} the results of this computation on a torus are shown. The smaller patches are generated with a smaller tool radius. Here the tool path deviates little from the parameter curve of the meridian circle, which is the path of widest stripe with tangents in the principal directions of the greater principal curvatures. The isophotic curves are the parallel circles. A larger tool radius shows a significantly bigger deviation from the steepest moving directions.

\begin{figure}
\psone{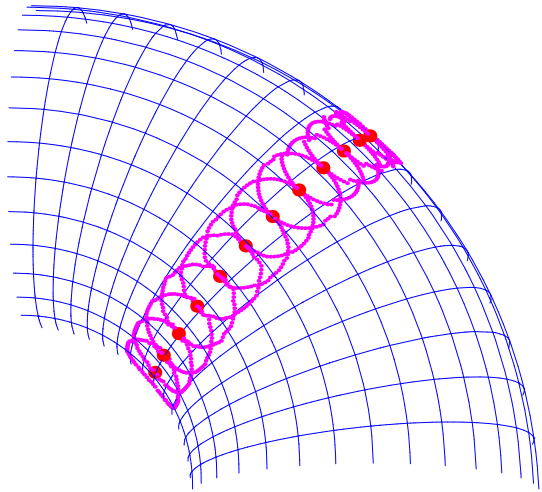}{8cm}
\caption{Tool path on the torus shown with the processed patches around the computed points.}\label{fig13}
\end{figure}

\begin{figure}
\psone{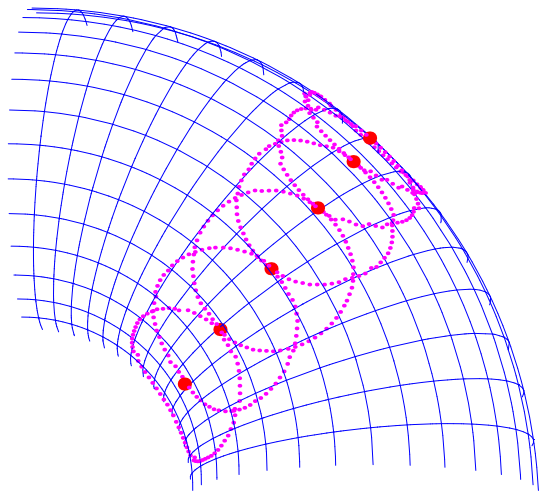}{8cm}
\caption{Tool path on the same torus with larger tool radius.}\label{fig14}
\end{figure}

The next surface in Figures \ref{fig15}, \ref{fig16} and \ref{fig17} is described by a trigonometric function. The steepest ascending direction crosses the bump (not shown), and an isophotic curve is shown in Figure \ref{fig15}. The computed tool path goes between them, and Figure \ref{fig16}, \ref{fig17} show that the computed moving directions are stable also in more curved regions.

\begin{figure}
\psone{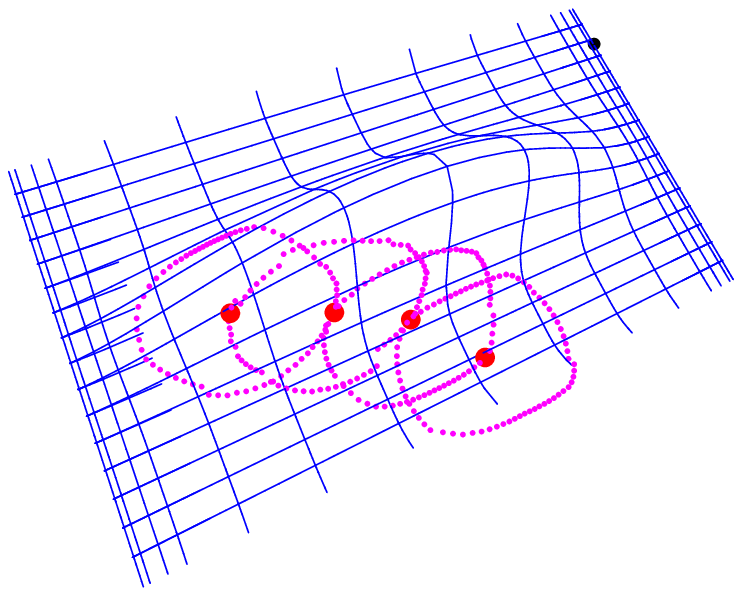}{8cm}
\caption{Isophotic curve  on a trigonometric surface.}\label{fig15}
\end{figure}

\begin{figure}
\psone{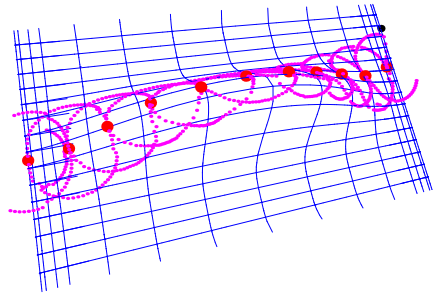}{8cm}
\caption{Tool paths in a more  curved region.}\label{fig16}
\end{figure}

\begin{figure}
\psone{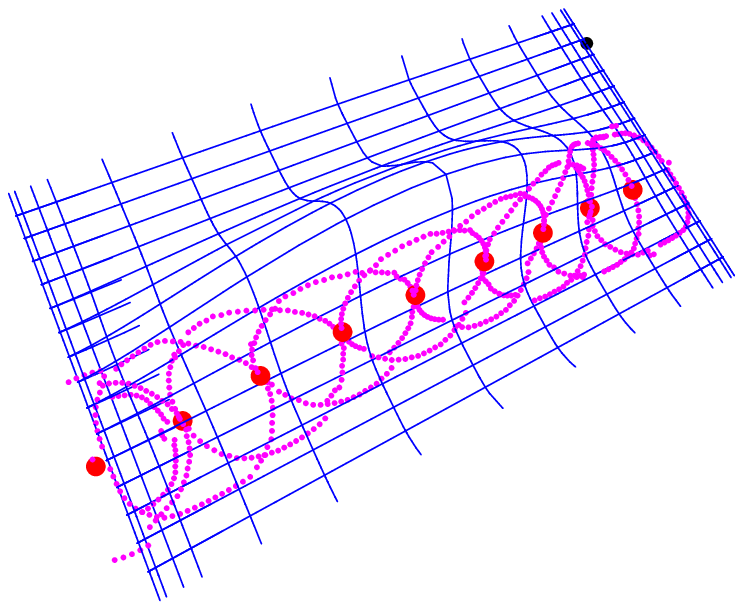}{8cm}
\caption{Tool paths in a  less curved region.}\label{fig17}
\end{figure}

The results in the examples correspond to expected ones, the direction choice optimizes the tool inclination angle for maximal material removal rate in each cutter contact point. The suggested locally optimal moving direction is depending as well on the surface curvatures at the actual point, as on the tool's size.

The analysis of the variation of the angle between the tool axis and the surface normal shows the following values. On the torus with Gaussian curvature between $-0.01$ and $0.08$ the change of this angle of the axis of the smaller tool (Fig. \ref{fig13}) along the path in the maximal curvature direction would be between $0.32$ and $0.4$, while along the path in the proposed direction between $0.28$ and $0.4$. The width of the processed patches has decreased by $\approx 0.16\%$. (Remember that this angle does not change when the tool is moving in the isophotic direction, but the change is maximal in the maximal normal curvature direction.) By our method in the case of the bigger tool (Fig. \ref{fig14}) the variation of the angle of the axis to the surface normal has been reduced to the interval $[0.1,0.34]$, and the width of the processed stripe has been reduced a bit more, by $\approx 0.3\%$. On the trigonometric surface with Gaussian curvature between $-0.001$ and $0.0006$ the variation of the observed angle along the steepest direction would be between $0.15$ and $0.44$, while along the proposed path between $0.0$ and $0.4$ (Fig. \ref{fig16}, \ref{fig17}). The width's decrease on the processed patches is in this case $2\%$. 

\section{Conclusions}
Geometric considerations about generating milling tool paths have been presented. Local moving direction of the tool has been computed, when the processed stripe is the widest, and the change of the inclination angle of the tool axis is minimal. A correction factor to the steepest ascending direction has been determined from the change of the isophotic angle of the surface normal with respect to the tool axis while moving the tool from a given point into the next position.
This strategy provides to develop a new  milling strategy with possible wide processed stripes and small variation of the angle between the tool axis and the surface normals. The presented new local offsetting method for triangular meshes solves basic problems, as computing the processed patch boundary and detecting self intersection or gauging.
For the solution of different other technical problems further investigations are necessary.

On analytical surfaces the computations and the figures have been made by the symbolic algebraic program package Mathematica.
The program developed by the author for the presented computations on discrete surfaces is not for professional applications. A lot of numerical problems may arise in the computation with real triangular meshes. The solution of such problems was not the task of this work.

\section{Acknowledgements}
The work was supported by a joint
project between the TU Berlin and the BUTE.

\end{JGGarticle}
\end{document}